# Local negative circuits and fixed points in Boolean networks

Adrien Richard

Laboratoire I3S, UMR 6070 CNRS & Université de Nice-Sophia Antipolis,
2000 route des Lucioles, 06903 Sophia Antipolis, France.

e-mail: `richard@i3s.unice.fr`
telephone: +33 4 92 94 27 51
fax: +33 4 92 94 28 98

**Abstract:** To each Boolean function $F : \{0,1\}^n \to \{0,1\}^n$ and each point $x \in \{0,1\}^n$, we associate the signed directed graph $G_F(x)$ of order $n$ that contains a positive (resp. negative) arc from $j$ to $i$ if the discrete analogue of $(\partial f_i/\partial x_j)(x)$ is positive (resp. negative). We then focus on the following open problem: Is the absence of a negative circuit in $G_F(x)$ for all $x \in \{0,1\}^n$ a sufficient condition for $F$ to have at least one fixed point? As main result, we settle this problem under the additional condition that, for all $x \in \{0,1\}^n$, the out-degree of each vertex of $G_F(x)$ is at most one.

# 1 Introduction

In the course of his analysis of discrete iterations, Robert introduced a discrete Jacobian matrix for Boolean maps and the notion of Boolean eigenvalue [2, 3, 4, 5]. This material allows Shih and Ho to state in 1999 a Boolean analogue of the Jacobian conjecture [7]: If a map from $\{0,1\}^n$ to itself is such that all the Boolean eigenvalues of the discrete Jacobian matrix of each element of $\{0,1\}^n$ are zero, then it has a unique fixed point. Thanks to the work of Shih and Dong [6], this conjecture is now a theorem.

Our starting point is an equivalent statement of the Shih-Dong theorem, the Theorem 1 below, in which the condition "all the Boolean eigenvalues of the discrete Jacobian matrix are zero" is expressed with the following few basic definitions and graph-theoretic notions.

Let $n$ be a positive integer, and consider a Boolean map

$$F : \{0,1\}^n \to \{0,1\}^n, \qquad x = (x_1, \ldots, x_n) \mapsto F(x) = (f_1(x), \ldots, f_n(x)).$$

The *interaction graph of $F$ evaluated at point $x \in \{0,1\}^n$* is the directed graph on $\{1, \ldots, n\}$ that contains an arc from a vertex $j$ to a vertex $i$ if the quantity

$$f_{ij}(x) = f_i(x_1, \ldots, x_{j-1}, 1, x_{j+1}, \ldots, x_n) - f_i(x_1, \ldots, x_{j-1}, 0, x_{j+1}, \ldots, x_n)$$

is not zero, *i.e.*, if the partial derivative of $f_i$ with respect to $x_j$ is not is not zero at point $x$. A *circuit* of length $p$ in $G_F(x)$ is a sequence of $p$ distinct vertices $i_1, i_2, \ldots, i_p$ such that there is an arc from $i_k$ to $i_{k+1}$, $1 \le k < p$, and from $i_p$ to $i_1$. An arc from a vertex to itself is thus a circuit of length one.

**Theorem 1 (Shih and Dong, 2005)**

If $G_F(x)$ has no circuit for all $x \in \{0,1\}^n$, then $F$ has a unique fixed point.

Remy, Ruet and Thieffry [1] proved latter that $F$ has at most one fixed point under a condition weaker than "$G_F(x)$ has no circuit for all $x \in \{0,1\}^n$". For that, they define the *sign of an arc* from $j$ to $i$ in $G_F(x)$ to be equals to $f_{ij}(x)$. And, as usual, they define the *sign of a circuit* to be the product of the signs of its edges.

**Theorem 2 (Remy, Ruet and Thieffry, 2008)**

If $G_F(x)$ has no positive circuit for all $x \in \{0,1\}^n$, then $F$ has at most one fixed point.

This theorem positively answer a Boolean version of a conjecture of Thomas coming from theoretical biology (see [1] and the references therein).

Seeing Theorems 1 and 2, it is natural to think about a proof by dichotomy of Theorem 1, and to study the following difficult question:

**Question 1** *Is the absence of a negative circuit in $G_F(x)$ for all $x \in \{0,1\}^n$ a sufficient condition for $F$ to have at least one fixed point?*

In this note, we partially answer this question by establishing the following theorem:

**Theorem 3** *If $G_F(x)$ has no negative circuit for all $x \in \{0,1\}^n$, and if the out-degree of each vertex of $G_F(x)$ is at most one for all $x \in \{0,1\}^n$, then $F$ has at least one fixed point.*

This partial answer is, in our knowledge, the first result about negative circuits in local interaction graphs associated with $F$. And it is not an obvious exercise. To see this, one can refer to the technical arguments used by Shih and Ho [7, pages 75-88] to prove that if $G_F(x)$ has no circuit for all $x \in \{0,1\}^n$, and if the out-degree of each vertex of $G_F(x)$ is at most one for all $x \in \{0,1\}^n$, then $F$ has at least one fixed point.

Finally, we also prove, using Theorem 2, the following theorem:

**Theorem 4** *If $G_F(x)$ has no negative circuit for all $x \in \{0,1\}^n$, and if there exists a vertex $i \in \{1,\ldots,n\}$ such that, for all $x \in \{0,1\}^n$, all the positive circuits of $G_F(x)$ contain $i$, then $F$ has at least one fixed point.*

Note that Theorem 1 is an immediate consequence of Theorem 2 and Theorem 4.

The paper is organized as follows. After some preliminaries given in Section 2, Sections 3 and 4 are devoted to the proof of Theorems 3 and 4 respectively.

## 2  Preliminaries

As usual, we set $\overline{0} = 1$ and $\overline{1} = 0$. For all $x \in \{0,1\}$ and $I \subseteq \{1,\ldots,n\}$, we denote by $\overline{x}^I$ the point $y$ of $\{0,1\}^n$ defined by: $y_i = \overline{x_i}$ if $i \in I$, and $y_i = x_i$ otherwise ($i = 1,\ldots,n$). In order to simplify notations, we write $\overline{x}$ instead of $\overline{x}^{\{1,\ldots,n\}}$, and $\overline{x}^i$ instead of $\overline{x}^{\{i\}}$.

Let $F$ be a map from $\{0,1\}^n$ to itself. Using the previous notations, the partial derivative of $f_i$ with respect to $x_j$ can be defined by

$$f_{ij}(x) = \frac{f_i(\overline{x}^j) - f_i(x)}{\overline{x_j} - x_j}.$$

If $G_F(x)$ has an arc from $j$ to $i$, we say that $i$ (resp. $j$) is a *successor* (resp. *predecessor*) of $j$ (resp. $i$), and we abusively write $j \to i \in G_F(x)$. The *out-degree* of a vertex is defined to be the number of successors of this vertex.

We are interested in maps $F$ that have the following property $\mathfrak{P}$:

$$\forall x \in \{0,1\}^n, \text{ the out-degree of each vertex of } G_F(x) \text{ is at most one.} \qquad (\mathfrak{P})$$

Note that if $F$ has the property $\mathfrak{P}$, then

$$j \to i \in G_F(x) \iff F(\overline{x}^i) = \overline{F(x)}^j.$$

The *Hamming distance* $d(x,y)$ between two points $x, y$ of $\{0,1\}^n$ is the number of indices $i \in \{1, \ldots, n\}$ such that $x_i \neq y_i$. So, for instance, $d(x,y) = n$ if and only if $y = \overline{x}$, and $d(x,y) = 1$ if and only if there exists $i \in \{1, \ldots, n\}$ such that $y = \overline{x}^i$. Note also that $F$ has the property $\mathfrak{P}$ if and only if

$$\forall x, y \in \{0,1\}^n, \qquad d(x,y) = 1 \implies d(F(x), F(y)) \leq 1.$$

We then deduce, by recurrence on $d(x,y)$, that $F$ has the property $\mathfrak{P}$ if and only if

$$\forall x, y \in \{0,1\}^n, \qquad d(F(x), F(y)) \leq d(x, y).$$

We now associate with $F$ two maps from $\{0,1\}^{n-1}$ to itself that will be used as inductive tools in the proof of Theorems 3 and 4. If $x \in \{0,1\}^{n-1}$ and $b \in \{0,1\}$, we denote by $(x, b)$ the point $(x_1, \ldots, x_{n-1}, b)$ of $\{0,1\}^n$. Then, for $b \in \{0,1\}$, we define the map $F^{|b} = (f_1^{|b}, \ldots, f_n^{|b}) : \{0,1\}^{n-1} \to \{0,1\}^{n-1}$ by

$$f_i^{|b}(x) = f_i(x, b) \qquad (i = 1, \ldots, n-1).$$

We have then the following obvious property: for all $x \in \{0,1\}^{n-1}$ and $b \in \{0,1\}$,

$$f_{ij}^{|b}(x) = f_{ij}(x,b) \qquad (i,j = 1, \ldots, n-1).$$

Consequently, for all $x \in \{0,1\}^{n-1}$ and $b \in \{0,1\}$,

$$G_{F^{|b}}(x) \text{ is a } subgraph \text{ of } G_F(x,b),$$

i.e., if $G_{F^{|b}}(x)$ has a positive (resp. negative) arc from $j$ to $i$, then $G_F(x,b)$ has a positive (resp. negative) arc from $j$ to $i$. It is then clear that if $F$ has the property $\mathfrak{P}$ then $F^{|b}$ has the property $\mathfrak{P}$.

## 3 Proof of Theorem 3

**Lemma 1** *If $d(x, F(x)) = 1$, then any circuit of $G_F(x)$ of length $n$ is negative.*

**Proof** – Suppose that $d(x, F(x)) = 1$ and that $C = i_1, \ldots, i_n$ is a circuit of $G_F(x)$ of length $n$. Without loss of generality, we can suppose that $F(x) = \overline{x}^{i_1}$. Let $h(1) = 1$ and $h(0) = -1$. We have

$$f_{i_1 i_n}(x) = \frac{f_{i_1}(\overline{x}^{i_n}) - f_{i_1}(x)}{\overline{x_{i_n}} - x_{i_n}} = \frac{f_{i_1}(\overline{x}^{i_n}) - \overline{x_{i_1}}}{\overline{x_{i_n}} - x_{i_n}},$$

and since $f_{i_1 i_n}(x) \neq 0$ we obtain

$$f_{i_1 i_n}(x) = \frac{x_{i_1} - \overline{x_{i_1}}}{\overline{x_{i_n}} - x_{i_n}} = \frac{h(x_{i_1})}{h(\overline{x_{i_n}})}.$$

Furthermore, for $k = 1, \ldots, n-1$, we have

$$f_{i_{k+1} i_k}(x) = \frac{f_{i_{k+1}}(\overline{x}^{i_k}) - f_{i_{k+1}}(x)}{\overline{x_{i_k}} - x_{i_k}} = \frac{f_{i_{k+1}}(\overline{x}^{i_k}) - x_{i_{k+1}}}{\overline{x_{i_k}} - x_{i_k}},$$

and since $f_{i_{k+1} i_k}(x) \neq 0$ we obtain

$$f_{i_{k+1} i_k}(x) = \frac{\overline{x_{i_{k+1}}} - x_{i_{k+1}}}{\overline{x_{i_k}} - x_{i_k}} = \frac{h(\overline{x_{i_{k+1}}})}{h(\overline{x_{i_k}})}.$$

Denoting by $s$ the sign of $C$, we obtain

$$\begin{aligned} s &= f_{i_2 i_1}(x) \cdot f_{i_3 i_2}(x) \cdot f_{i_4 i_3}(x) \cdots f_{i_n i_{n-1}}(x) \cdot f_{i_1 i_n}(x) \\ &= \frac{h(\overline{x_{i_2}})}{h(\overline{x_{i_1}})} \cdot \frac{h(\overline{x_{i_3}})}{h(\overline{x_{i_2}})} \cdot \frac{h(\overline{x_{i_4}})}{h(\overline{x_{i_3}})} \cdots \frac{h(\overline{x_{i_n}})}{h(\overline{x_{i_{n-1}}})} \cdot \frac{h(x_{i_1})}{h(\overline{x_{i_n}})} = \frac{h(x_{i_1})}{h(\overline{x_{i_1}})} = -1. \end{aligned}$$

$\square$

The rest of the proof is based on the following notion of opposition: given two points $x, y \in \{0, 1\}^n$ and an index $i \in \{1, \ldots, n\}$, we say that $x$ and $y$ are in *opposition* (*with respect to $i$ in $F$*) if

$$F(x) = \overline{x}^i, \qquad F(y) = \overline{y}^i \qquad \text{and} \qquad x_i \neq y_i.$$

**Lemma 2** *Let $F$ be a map from $\{0, 1\}^n$ to itself that has the property $\mathfrak{P}$. If $F$ has two points in opposition, then there exists two distinct points $x$ and $y$ in $\{0, 1\}^n$ such that $G_F(x)$ and $G_F(y)$ have a common negative circuit.*

**Proof** – We proceed by induction on $n$. The lemma being obvious for $n = 1$, we suppose that $n > 1$ and that the lemma holds for maps from $\{0, 1\}^{n-1}$ to itself. We also suppose that $F$ has at least two points in opposition.

First, suppose that $\alpha$ and $\beta$ are two points in opposition with respect to $i$ in $F$ such that $\alpha \neq \overline{\beta}$. Then there exists $j \neq i$ such that $\alpha_j = \beta_j$, and without loss of generality we can suppose that $\alpha_n = \beta_n = b$. Set $\tilde{\alpha} = (\alpha_1, \ldots, \alpha_{n-1})$ and $\tilde{\beta} = (\beta_1, \ldots, \beta_{n-1})$ so that $\alpha = (\tilde{\alpha}, b)$ and $\beta = (\tilde{\beta}, b)$. Then, $\tilde{\alpha}_i = \alpha_i \neq \beta_i = \tilde{\beta}_i$, and since $F(\alpha) = \overline{\alpha}^i$, we have

$$F^{|b}(\tilde{\alpha}) = (f_1(\alpha), \ldots, f_i(\alpha), \ldots, f_{n-1}(\alpha)) = (\alpha_1, \ldots, \overline{\alpha_i}, \ldots, \alpha_{n-1}) = \overline{\tilde{\alpha}}^i,$$

and we show similarly that $F^{|b}(\tilde{\beta}) = \overline{\tilde{\beta}}^i$. Consequently, $\tilde{\alpha}$ and $\tilde{\beta}$ are in opposition with respect to $i$ in $F^{|b}$. Since $F$ has the property $\mathfrak{P}$, $F^{|b}$ has the property $\mathfrak{P}$, and so, by induction hypothesis, there exists two distinct points $x, y \in \{0,1\}^{n-1}$ such that $G_{F^{|b}}(x)$ and $G_{F^{|b}}(y)$ have a common negative circuit. Since $G_{F^{|b}}(x)$ and $G_{F^{|b}}(y)$ are subgraphs of $G_F(x, b)$ and $G_F(y, b)$ respectively, we deduce that $G_F(x, b)$ and $G_F(y, b)$ have a common negative circuit and the lemma holds.

So in the following, we assume the following hypothesis $\mathfrak{H}$:

$$\text{If } F \text{ has two points } \alpha \text{ and } \beta \text{ in opposition, then } \alpha = \overline{\beta}. \tag{$\mathfrak{H}$}$$

We need the following four claims to complet the proof.

**Claim 1** *F has no fixed point.*

*Proof* – Let $\alpha$ and $\beta$ be two points in opposition with respect to $i$ in $F$. Suppose, by contradiction, that $x$ is a fixed point of $F$. If $x_i = \alpha_i$, then $d(F(x), F(\alpha)) = d(x, \overline{\alpha}^i) > d(x, \alpha)$ and this contradicts the fact that $F$ has the property $\mathfrak{P}$. Otherwise, $x_i = \beta_i$, thus

$d(F(x), F(\beta)) = d(x, \overline{\beta}^i) > d(x, \beta)$ and we arrive to the same contradiction. □

**Notation:** *In the following, for all $x \in \{0,1\}^n$, we set*

$$x^1 = x \quad \text{and} \quad x^{k+1} = F(x^k) \quad (k = 1, 2, 3, \dots).$$

**Claim 2** *If $\alpha$ and $\beta$ are in opposition in $F$, then there exists a permutation $\{i_1, \dots, i_n\}$ of $\{1, \dots, n\}$ such that $\alpha^k$ and $\beta^k$ are in opposition with respect to $i_k$ in $F$ $(k = 1, \dots, n)$.*

*Proof* – Suppose that $\alpha = \alpha^1$ and $\beta = \beta^1$ are in opposition with respect to $i$ in $F$. For $p = 1, \dots, n$, we denote by $S_p$ the set of sequences $(i_1, i_2, \dots, i_p)$ of $p$ distinct indices of $\{1, \dots, n\}$ such that $\alpha^{k+1} = \overline{\alpha^k}^{i_k}$ for $k = 1, \dots, p$. $S_1$ is not empty since, by definition, $(i) \in S_1$. So in order to prove that $S_n$ is not empty, it is sufficient to prove that

$$S_p \neq \emptyset \Rightarrow S_{p+1} \neq \emptyset \quad (p = 1, \dots, n-1).$$

Suppose that $(i_1, \dots, i_p) \in S_p$ $(1 \leq p < n)$. Then $\alpha^{p+1} = \overline{\alpha^p}^{i_p}$, so $d(\alpha^{p+1}, \alpha^p) = 1$ and since $F$ has the property $\mathfrak{P}$, we deduce that

$$d(F(\alpha^{p+1}), \alpha^{p+1}) = d(F(\alpha^{p+1}), F(\alpha^p)) \leq d(\alpha^{p+1}, \alpha^p) = 1.$$

Since, by Claim 1, we have $F(\alpha^{p+1}) \neq \alpha^{p+1}$, we deduce that $d(F(\alpha^{p+1}), \alpha^{p+1}) = 1$. In other words, there exists $j \in \{1, \dots, n\}$ such that

$$F(\alpha^{p+1}) = \overline{\alpha^{p+1}}^j.$$

Suppose that there exists $k \in \{1, \ldots, p\}$ such that $j = i_k$. Then,

$$F(\alpha^k) = \overline{\alpha^k}^j$$

and since

$$\alpha^{p+1} = \overline{\alpha^p}^{\{i_p\}} = \overline{\alpha^{p-1}}^{\{i_{p-1}, i_p\}} = \cdots = \overline{\alpha^k}^{\{i_k, \ldots, i_{p-1}, i_p\}},$$

we have

$$\alpha_j^k = \alpha_{i_k}^k \neq \alpha_{i_k}^{p+1} = \alpha_j^{p+1}.$$

Thus $\alpha^k$ and $\alpha^{p+1}$ are in opposition with respect to $i$ in $F$. But since $\{i_k, \ldots, i_{p-1}, i_p\}$ is strictly included in $\{1, \ldots, n\}$, we have $\alpha^{p+1} \neq \overline{\alpha^k}$ and this contradicts the hypothesis $\mathfrak{H}$. Thus $j \notin \{i_1, \ldots, i_p\}$ and we deduce that $(i_1, \ldots, i_p, j)$ belongs to $S_{p+1}$. Thus $S_{p+1}$ is not empty and it follows that $S_n$ is not empty. Thus, there exists a permutation $\{i_1, \ldots, i_n\}$ of $\{1, \ldots, n\}$ such that $\alpha^{p+1} = \overline{\alpha^p}^{i_p}$ for $p = 1, \ldots, n$, and we show similarly that there exists a permutation $\{j_1, \ldots, j_n\}$ of $\{1, \ldots, n\}$ such that $\beta^{p+1} = \overline{\beta^p}^{j_p}$ for $p = 1, \ldots, n$. Observe that, following the hypothesis $\mathfrak{H}$, we have $\alpha = \overline{\beta}$ and thus

$$\alpha^{n+1} = \overline{\alpha}^{\{i_1, \ldots, i_n\}} = \overline{\alpha} = \beta \quad \text{and} \quad \beta^{n+1} = \overline{\beta}^{\{j_1, \ldots, j_n\}} = \overline{\beta} = \alpha. \tag{1}$$

We are now in possition to prove, by recurrence on $k$ decreasing from $n$ to $1$, that $\alpha^k$ and $\beta^k$ are in opposition with respect to $i_k$ in $F$. Since $F$ has the property $\mathfrak{P}$, and from (1), we have

$$d(\alpha^n, \beta^n) \geq d(F(\alpha^n), F(\beta^n)) = d(\alpha^{n+1}, \beta^{n+1}) = d(\beta, \alpha) = d(\beta, \overline{\beta}) = n.$$

thus
$$d(\alpha^n, \beta^n) = n = d(\alpha^{n+1}, \beta^{n+1}) = d(\overline{\alpha^n}^{i_n}, \overline{\beta^n}^{j_n})$$

We deduce that $i_n = j_n$ and $\alpha_{i_n}^n \neq \beta_{i_n}^n$. It is then clear that $\alpha^n$ and $\beta^n$ are in opposition with respect to $i_n$ in $F$. Now, suppose that $\alpha^k$ and $\beta^k$ are in opposition with respect to $i_k$ in $F$ ($2 \leq k \leq n$). Then, following the hypothesis $\mathfrak{H}$, $\alpha^k = \overline{\beta^k}$, and since $F$ has the property $\mathfrak{P}$, we deduce that

$$d(\alpha^{k-1}, \beta^{k-1}) \geq d(F(\alpha^{k-1}), F(\beta^{k-1})) = d(\alpha^k, \beta^k) = d(\overline{\beta^k}, \beta^k) = n$$

Thus
$$d(\alpha^{k-1}, \beta^{k-1}) = n = d(\alpha^k, \beta^k) = d(\overline{\alpha^{k-1}}^{i_{k-1}}, \overline{\beta^{k-1}}^{j_{k-1}}).$$

We deduce that $i_{k-1} = j_{k-1}$ and $\alpha_{i_{k-1}}^{k-1} \neq \alpha_{i_{k-1}}^{k-1}$ and thus that $\alpha^{k-1}$ and $\beta^{k-1}$ are in opposition with respect to $i_{k-1}$ in $F$. $\square$

**Claim 3** *If $\alpha$ and $\beta$ are in opposition with respect to $i$ in $F$, then $i$ has at most one predecessor in $G_F(\alpha)$.*

*Proof* – Let $\{i_1, \ldots, i_n\}$ be a permutation of $\{1, \ldots, n\}$ with the property of Claim 2. Then $\overline{\alpha}^{i_1} = F(\alpha) = \overline{\alpha}^i$ thus $i = i_1$. Suppose, by contradiction, that $i_1$ has at least two predecessors in $G_F(\alpha)$. Then $i_1$ has a predecessor $i_k \neq i_n$ in $G_F(\alpha)$. Using the property $\mathfrak{P}$, we deduce that

$$F(\overline{\alpha}^{i_k}) = \overline{F(\alpha)}^{i_1} = \overline{\overline{\alpha}^{i_1}}^{i_1} = \alpha = \overline{\overline{\alpha}^{i_k}}^{i_k} \quad \text{and} \quad F(\alpha^k) = \overline{\alpha^k}^{i_k}.$$

If $k = 1$, then $\alpha^k = \alpha$ and so

$$(\alpha^k)_{i_k} = \alpha_{i_k} \neq (\overline{\alpha}^{i_k})_{i_k} \quad \text{and} \quad \alpha^k_{i_n} = (\overline{\alpha}^{i_k})_{i_n}. \tag{2}$$

Otherwise, $\alpha^k = \overline{\alpha}^{\{i_1,\ldots,i_{k-1}\}}$ and so (2) holds again. Consequently, in both cases, $\alpha^k$ and $\overline{\alpha}^{i_k}$ are in opposition with respect to $i_k$ in $F$ and $\alpha^k \neq \overline{\alpha}^{i_k}$. This contradicts the hypothesis $\mathfrak{H}$. □

**Claim 4** *If $\alpha$ et $\beta$ are in opposition in $F$, then $G_F(\alpha^n)$ has a circuit of length $n$.*

*Proof* – Let $\{i_1, \ldots, i_n\}$ be a permutation of $\{1, \ldots, n\}$ with the property of Claim 2. We will show that $i_1, \ldots, i_n$ is a circuit of $G_F(\alpha^n)$. We have

$$F\left(\overline{\alpha^k}^{i_{k-1}}\right) = F\left(\overline{\overline{\alpha^{k-1}}^{i_{k-1}}}^{i_{k-1}}\right) = F(\alpha^{k-1}) = \alpha^k = \overline{\overline{\alpha^k}^{i_k}}^{i_k} = \overline{F(\alpha^k)}^{i_k} \quad (k = 2, \ldots, n)$$

and thus

$$i_{k-1} \to i_k \in G_F(\alpha^k) \quad (k = 2, \ldots, n). \tag{3}$$

In addition,

$$F\left(\overline{\alpha^k}^{i_k}\right) = F(\alpha^{k+1}) = \overline{\alpha^{k+1}}^{i_{k+1}} = \overline{F(\alpha^k)}^{i_{k+1}} \quad (k = 1, \ldots, n-1)$$

and thus

$$i_k \to i_{k+1} \in G_F(\alpha^k) \quad (k = 1, \ldots, n-1).$$

Let $k$ be any index of $\{1, \ldots, n-1\}$, and suppose, by contradiction, that

$$i_k \to i_{k+1} \notin G_F(\alpha^n).$$

Since $i_k \to i_{k+1} \in G_F(\alpha^k)$, there exists $p \in \{k+1, \ldots, n\}$ such that

$$i_k \to i_{k+1} \in G_F(\alpha^{p-1}) \quad \text{and} \quad i_k \to i_{k+1} \notin G_F(\alpha^p).$$

Following (3), we have $i_p \neq i_{k+1}$. Furthermore, from $i_k \to i_{k+1} \in G_F(\alpha^{p-1})$ we deduce that

$$f_{i_{k+1}}(\alpha^{p-1}) \neq f_{i_{k+1}}\big(\overline{\alpha^{p-1}}^{i_k}\big), \tag{4}$$

and from both $i_k \to i_{k+1} \notin G_F(\alpha^p)$ and $\alpha^p = \overline{\alpha^{p-1}}^{i_{p-1}}$ we deduce that

$$f_{i_{k+1}}\big(\overline{\alpha^{p-1}}^{i_{p-1}}\big) = f_{i_{k+1}}\left(\overline{\overline{\alpha^{p-1}}^{i_{p-1}}}^{i_k}\right) = f_{i_{k+1}}\left(\overline{\overline{\alpha^{p-1}}^{i_k}}^{i_{p-1}}\right). \tag{5}$$

If

$$f_{i_{k+1}}(\alpha^{p-1}) \neq f_{i_{k+1}}\big(\overline{\alpha^{p-1}}^{i_{p-1}}\big)$$

then $i_{k+1}$ and $i_p$ are distinct successors of $i_{p-1}$ in $G_F(\alpha^{p-1})$, and this contradicts the fact that $F$ has the property $\mathfrak{P}$. Thus

$$f_{i_{k+1}}(\alpha^{p-1}) = f_{i_{k+1}}\big(\overline{\alpha^{p-1}}^{i_{p-1}}\big)$$

and from (4) and (5) we deduce that

$$f_{i_{k+1}}(\overline{\alpha^{p-1}}^{i_k}) \neq f_{i_{k+1}}\left(\overline{\overline{\alpha^{p-1}}^{i_k}}^{i_{p-1}}\right).$$

Thus $i_{p-1} \to i_{k+1} \in G_F(\overline{\alpha^{p-1}}^{i_k})$ and since $F$ has the property $\mathfrak{P}$, we have

$$F(\overline{\alpha^p}^{i_k}) = F\left(\overline{\overline{\alpha^{p-1}}^{i_{p-1}}}^{i_k}\right) = F\left(\overline{\overline{\alpha^{p-1}}^{i_k}}^{i_{p-1}}\right) = \overline{F(\overline{\alpha^{p-1}}^{i_k})}^{i_{k+1}}$$

Since $i_k \to i_{k+1} \in G_F(\alpha^{p-1})$, we have $F(\overline{\alpha^{p-1}}^{i_k}) = \overline{F(\alpha^{p-1})}^{i_{k+1}}$ and using the property $\mathfrak{P}$ we obtain

$$F(\overline{\alpha^p}^{i_k}) = \overline{\overline{F(\alpha^{p-1})}^{i_{k+1}}}^{i_{k+1}} = F(\alpha^{p-1}) = \alpha^p = \overline{\alpha^p}^{i_p} = \overline{F(\alpha^p)}^{i_p}$$

So $i_k$ and $i_{p-1}$ are predecessors of $i_p$ in $G_F(\alpha^p)$, and $i_k \neq i_{p-1}$ since $i_p \neq i_{k+1}$. We have now a contradiction: following Claim 2, $\alpha^p$ and $\beta^p$ are in opposition with respect to $i_p$ in $F$, and so, following Claim 3, $i_p$ has at most one predecessor in $G_F(\alpha^p)$. We have thus prove that

$$i_k \to i_{k+1} \in G_F(\alpha^n) \qquad (k = 1, \ldots, n-1)$$

To prove the claim, it is thus sufficient to prove that $i_n \to i_1 \in G_F(\alpha^n)$, and this is obvious. Indeed, following the hypothesis $\mathfrak{H}$, we have $\overline{\alpha} = \beta$, thus

$$F(\alpha^n) = \alpha^{n+1} = \overline{\alpha}^{\{i_1,\ldots,i_n\}} = \overline{\alpha} = \beta$$

and so

$$F\bigl(\overline{\alpha^n}^{i_n}\bigr) = F(\alpha^{n+1}) = F(\beta) = \overline{\beta}^{i_1} = \overline{F(\alpha^n)}^{i_1}.$$

□

We are now in position to prove the lemma. Let $\alpha$ and $\beta$ be two points in opposition in $F$. Following Claim 2 and Claim 4, $\alpha^n$ and $\beta^n$ are two points in opposition, and thus distinct, such that $G_F(\alpha^n)$ and $G_F(\beta^n)$ have a common circuit of length $n$, and according to Lemma 1, this circuit is negative, both in $G_F(\alpha^n)$ and $G_F(\beta^n)$. □

**Lemma 3** *Let $F$ be a map from $\{0,1\}^n$ to itself that has the property $\mathfrak{P}$. If there is no distinct points $x, y \in \{0,1\}^n$ such that $G_F(x)$ and $G_F(y)$ have a common negative circuit, then $F$ has at least one fixed point.*

**Proof** – We proceed by induction on $n$. The lemma being obvious for $n = 1$, we suppose that $n > 1$ and that the lemma holds for maps from $\{0,1\}^{n-1}$ to itself. Let $F$ be as in the statement, and let $b \in \{0,1\}$. Since $G_{F^{|b}}(x)$ is a subgraph of $G_F(x, b)$ for all $x \in \{0,1\}^{n-1}$, $F^{|b}$ has the property $\mathfrak{P}$ and there is no distinct points $x, y \in \{0,1\}^n$ such that $G_{F^{|b}}(x)$ and $G_{F^{|b}}(y)$ have a common negative circuit. So, by induction hypothesis, $F^{|b}$ has at least one fixed point that we denote by $\xi^b$. Now, we prove that $(\xi^0, 0)$ or $(\xi^1, 1)$ is a fixed point of

$F$. If not, then for $b \in \{0,1\}$,

$$
\begin{aligned}
F(\xi^b, b) &= (f_1(\xi^b, b), \ldots, f_{n-1}(\xi^b, b), f_n(\xi^b, b)) \\
&= (f_1^{|b}(\xi^b), \ldots, f_{n-1}^{|b}(\xi^b), f_n(\xi^b, b)) \\
&= (\xi_1^b, \ldots, \xi_{n-1}^b, f_n(\xi^b, b)) \\
&= (\xi^b, f_n(\xi^b, b)) \\
&= (\xi^b, \bar{b}) \\
&= \overline{(\xi^b, b)}^n.
\end{aligned}
$$

We deduce that $(\xi^0, 0)$ and $(\xi^1, 1)$ are in opposition with respect to $n$ in $F$, and so, by Lemma 2, there exists two distinct points $x, y \in \{0,1\}^n$ such that $G_F(x)$ and $G_F(y)$ have a common negative circuit, a contradiction. $\square$

Theorem 1 is an obvious consequence of Lemma 3.

## 4 Proof of Theorem 4

We proceed by induction on $n$. The case $n = 1$ being obvious, we suppose that $n > 1$ and that the theorem holds for maps from $\{0,1\}^{n-1}$ to itself. Let $F$ be a map from $\{0,1\}^n$ to itself, and without loss of generality, suppose that, for all $x \in \{0,1\}^n$, all the positive circuits of $G_F(x)$ contain the vertex $n$.

For $b \in \{0,1\}$ and $x \in \{0,1\}^{n-1}$, it is clear that $G_{F^{|b}}(x)$ has no circuit since $G_{F^{|b}}(x)$ is a subgraph of $G_F(x, b)$ that does not contains the vertex $n$. So $F^{|b}$ trivially satisfies the conditions of the theorem. So, by induction hypothesis, $F^{|b}$ has at least one fixed point that we denote by $\xi^b$.

We will show that $\alpha = (\xi^0, 0)$ or $\beta = (\xi^b, 1)$ is a fixed point of $F$. Suppose, by contradiction, that neither $\alpha$ nor $\beta$ is a fixed point of $F$. Then, as in Lemma 3, we prove that $F(\alpha) = \overline{\alpha}^n$ and that $F(\beta) = \overline{\beta}^n$.

Consider the map $\bar{F}$ from $\{0,1\}^n$ to $\{0,1\}^n$ defined by

$$\bar{F}(x) = \overline{F(x)}^n.$$

It is clear that $\alpha$ and $\beta$ are distinct fixed points of $\bar{F}$. So, by Theorem 2, there exists $x \in \{0,1\}^n$ such that $G_{\bar{F}}(x)$ has a positive circuit $C$. If $n$ does not belong to $C$, then since

$$\bar{f}_{ij} = f_{ij} \quad \text{for } i = 1, \ldots, n-1 \text{ and } j = 1, \ldots, n, \tag{6}$$

we deduce that $C$ is a positive circuit of $G_F(x)$ that does not contains $n$, a contradiction. Otherwise, $n$ belongs to $C$, and we then deduce from (6) and the fact that

$$\bar{f}_{nj} = -f_{nj} \quad \text{for } j = 1, \ldots, n$$

that $C$ is a negative circuit of $G_F(x)$, a contradiction.